\documentclass[11pt]{article}

\usepackage{amsmath, amssymb, amsfonts}
\usepackage{graphicx}
\usepackage{booktabs}
\usepackage{natbib}
\usepackage{hyperref}
\usepackage{geometry}
\title{Efficiency versus Robustness under Tail Misspecification:\\
Importance Sampling and Moment-Based VaR Bracketing
}

\author{
Aditri \\ 
Rutgers Business School \\
\texttt{aditri.aditri@rutgers.edu}
}

\date{\today}

\begin{document}
\maketitle

\begin{abstract}
Value-at-Risk (VaR) estimation at high confidence levels is inherently a rare-event problem and is particularly sensitive to tail behavior and model misspecification. This paper studies the performance of two simulation-based VaR estimation approaches, importance sampling (IS) and discrete moment matching (DMM), under controlled tail misspecification. The analysis separates the nominal model used for estimator construction from the true data-generating process used for evaluation, allowing the effects of heavy-tailed returns to be examined in a transparent and reproducible setting.

Daily returns of a broad equity market proxy are used to calibrate a nominal Gaussian model, while the true returns are generated from Student-$t$ distributions with varying degrees of freedom to represent increasingly heavy tails. Importance sampling is implemented via exponential tilting of the Gaussian model and VaR is estimated through likelihood-weighted root-finding. Discrete moment matching constructs deterministic lower and upper VaR bounds by enforcing a finite number of moment constraints on a discretized loss distribution.

The results demonstrate a clear trade-off between efficiency and robustness. Importance sampling produces low-variance VaR estimates under the nominal model but systematically underestimates the true VaR under heavy-tailed returns, with bias increasing at higher confidence levels and for thicker tails. Stability diagnostics such as effective sample size and weight concentration deteriorate as the estimation problem becomes more extreme. In contrast, DMM yields conservative VaR bracketing that remains robust under tail misspecification, with bounds tightening as additional moments are enforced up to a numerically feasible limit.

These findings highlight that variance reduction alone is insufficient for reliable tail risk estimation when model uncertainty is significant. Moment-based approaches provide a complementary framework by explicitly accounting for distributional ambiguity, offering robustness at the expense of efficiency. The paper provides practical diagnostic guidance for assessing the reliability of simulation-based VaR estimators under heavy-tailed market conditions.
\end{abstract}

\vspace{1em}


\newpage

\section{Introduction}
\label{sec:introduction}

Value-at-Risk (VaR) is one of the most widely used risk measures in financial risk management,
serving as both a regulatory benchmark and a practical tool for quantifying potential losses over
fixed horizons. At high confidence levels such as $99\%$ and above, VaR estimation becomes an
inherently rare-event problem: the target quantile is determined by extreme tail probabilities of
the loss distribution. From a computational standpoint, standard Monte Carlo methods can be
inefficient in this regime because only a small fraction of simulated samples fall into the relevant
tail region and therefore contribute meaningfully to the estimate.

Two broad methodological responses to this challenge motivate the present study.
The first focuses on \emph{computational efficiency} through variance-reduction techniques, among
which importance sampling (IS) plays a central role. By changing the underlying probability
measure to make tail events occur more frequently and correcting via likelihood ratios, IS can
substantially reduce estimator variance when the change of measure is well aligned with the
structure of extreme losses. This viewpoint is emphasized in the rare-event simulation literature,
notably in the work of Glasserman, Heidelberger, and Shahabuddin, where exponential tilting is
used to efficiently estimate rare-event probabilities relevant for portfolio risk.

A second response emphasizes \emph{robustness to distributional uncertainty} rather than sampling
efficiency. In many financial applications, the primary difficulty is not only Monte Carlo noise,
but also uncertainty about the correct tail model. It is well known that Gaussian return models
can be unrealistic for equity returns due to heavy tails, skewness, and volatility clustering; in such
settings, simulation can produce highly \emph{precise} estimates of VaR under a misspecified model.
Discrete Moment Matching (DMM) belongs to a class of moment-based approaches that weaken
parametric assumptions: given a finite set of moment constraints, DMM constructs a discrete
distributional approximation and, more fundamentally in the formulation adopted here, produces
\emph{bounds} on VaR over the set of all distributions consistent with the enforced moments and a
chosen discretization grid. Rather than returning a single number, DMM yields an interval that
explicitly reflects distributional ambiguity.

These perspectives highlight a key point: importance sampling and moment-based methods do not
solve the same problem. Importance sampling is designed to reduce variance when estimating tail
probabilities \emph{under a specified nominal model}, whereas moment-based procedures are designed
to quantify tail risk \emph{under partial distributional information}. This distinction becomes
particularly important under tail misspecification. If the nominal return model understates tail
thickness, then increasing the Monte Carlo budget or applying variance reduction can reduce
sampling error without correcting the underlying model error. In this sense, IS should be viewed
primarily as a computational efficiency enhancement rather than a remedy for model misspecification.

The objective of this paper is to provide a transparent comparison between these two approaches
in a controlled setting where tail misspecification is introduced deliberately and systematically.
We separate the \emph{nominal} model used for estimator construction from the \emph{true}
data-generating process used for evaluation. Daily returns of a broad equity market proxy are used
to calibrate a nominal Gaussian model, while true returns are generated from Student-$t$
distributions with varying degrees of freedom to represent increasingly heavy tails. Within this
framework, IS is implemented via exponential tilting of the Gaussian model and VaR is computed
through likelihood-weighted root-finding, while DMM constructs moment-feasible VaR bounds by
enforcing finitely many moment constraints on a discretized loss distribution. This design isolates
the efficiency-robustness trade-off: IS can be highly efficient for the nominal target yet systematically
underestimate tail risk under heavy-tailed behavior, whereas DMM provides conservative bracketing
that remains meaningful under tail uncertainty at the expense of informativeness and numerical
feasibility at high moment orders.

\paragraph{Contributions.}
This paper makes four contributions.
First, it provides a clear estimand-level comparison by distinguishing point estimation of VaR under
a nominal parametric model (IS) from interval-valued VaR characterization under moment-based
distributional ambiguity (DMM). Second, using a controlled simulation design that separates the
nominal Gaussian calibration from a heavy-tailed Student-$t$ data-generating process, it quantifies
how importance sampling remains computationally efficient while converging to the nominal-model
VaR, producing systematic underestimation of true tail risk as tails thicken and confidence levels
increase. Third, it empirically characterizes the behavior of DMM VaR bounds as additional moments
are enforced, documenting contraction of the moment-feasible set up to a numerically feasible limit
and interpreting resulting interval widths as an explicit measure of ambiguity rather than noise.
Fourth, it complements point estimates with practical diagnostics, including effective sample size
and weight concentration for IS and feasibility limits for DMM, that help assess numerical reliability
and the degree of misalignment between sampling measures, tail geometry, and distributional assumptions.

The remainder of the paper is organized as follows. Section~2 introduces the risk measure and loss
framework. Section~3 describes the data calibration and nominal return model. Section~4 presents
the heavy-tailed data-generating process used in the simulation study. Section~5 outlines the DMM
and IS VaR procedures. Section~6 describes the simulation design. Section~7 reports the empirical
findings, and Sections~8 to 9 discuss implications and conclude.

\section{Problem Setup and Risk Measure}
\label{sec:setup}

\subsection{Data and Return Construction}
Let $\{S_t\}_{t=0}^{T}$ denote the observed daily closing prices of QQQ over a fixed historical window, where $t$ indexes trading days. We define the one-day log-return by
\begin{equation}
R_t \;:=\; \log\!\left(\frac{S_t}{S_{t-1}}\right), 
\qquad t=1,\dots,T.
\label{eq:log_return_def}
\end{equation}
We treat $\{R_t\}_{t=1}^T$ as the calibration sample used to estimate parameters of a nominal return model.

\subsection{Loss Variable}
We consider a one-day risk horizon and define the loss random variable $L$ by
\begin{equation}
L \;:=\; -R,
\label{eq:loss_def}
\end{equation}
where $R$ is a generic one-day return distributed according to a model specified below. Under \eqref{eq:loss_def}, large losses correspond to large negative returns. All results in this paper are stated in ``return units'' (i.e., losses measured as negative log-returns). A dollar P\&L loss can be recovered by a monotone transformation if needed, but is not required for the comparative study considered here.

\subsection{Value-at-Risk as a Tail Probability Root}
Fix a confidence level $\alpha \in (0,1)$ (e.g., $\alpha=0.99$). The Value-at-Risk at level $\alpha$ is defined as the $\alpha$-quantile of the loss distribution:
\begin{equation}
\mathrm{VaR}_\alpha(L)
\;:=\;
\inf\Big\{x \in \mathbb{R} : F_L(x) \ge \alpha\Big\},
\qquad
F_L(x) := \mathbb{P}(L \le x).
\label{eq:var_def}
\end{equation}
Equivalently, letting
\begin{equation}
p(x) \;:=\; \mathbb{P}(L > x),
\label{eq:tail_prob_def}
\end{equation}
VaR is characterized as the solution of the tail probability equation
\begin{equation}
p(x_\alpha) \;=\; 1-\alpha,
\qquad \text{where } x_\alpha := \mathrm{VaR}_\alpha(L).
\label{eq:var_root_def}
\end{equation}
In the present setting $L=-R$, so the rare-event set $\{L>x\}$ can be expressed in terms of returns as
\begin{equation}
\{L>x\} \;=\; \{-R>x\} \;=\; \{R<-x\}.
\label{eq:event_equiv}
\end{equation}
Thus, VaR estimation at high confidence levels reduces to accurately estimating small tail probabilities of the form $\mathbb{P}(R<-x)$.

\subsection{Objective of the Study}
The goal of this paper is to compare two simulation-based approaches to
characterizing tail risk at level $\alpha$, centered around the Value-at-Risk
quantity $x_\alpha$ defined in \eqref{eq:var_root_def}:
(i) an importance sampling estimator based on an exponential change of measure applied to a Gaussian nominal model, and
(ii) a discrete moment matching estimator based on constructing a discrete approximation to the return distribution.
The comparison is performed under controlled tail misspecification by sampling returns from a heavy-tailed Student-$t$ data-generating process while applying estimators calibrated under a Gaussian nominal model. In short, while importance sampling targets a point estimate of $x_\alpha$ under the
nominal model, discrete moment matching produces interval-valued bounds that
reflect distributional ambiguity.

\section{Nominal Model and Calibration}
\label{sec:calibration}

\subsection{Nominal Probability Measure}
Throughout this paper, VaR estimators are constructed under a \emph{nominal} probability measure, denoted by $\mathbb{P}$, which serves as the assumed return model for both importance sampling and discrete moment matching. Under $\mathbb{P}$, the one-day return $R$ is modeled as Gaussian:
\begin{equation}
R \sim \mathcal{N}(\mu,\sigma^2),
\label{eq:nominal_gaussian}
\end{equation}
where $(\mu,\sigma)$ are unknown parameters to be estimated from historical data. This Gaussian model is not intended to be a realistic description of extreme tail behavior; rather, it provides a tractable baseline that allows the effect of variance-reduction and moment-matching techniques to be studied in isolation.

\subsection{Parameter Estimation}
Let $\{R_t\}_{t=1}^T$ denote the observed log-return sample defined in \eqref{eq:log_return_def}. The nominal parameters $(\mu,\sigma)$ in \eqref{eq:nominal_gaussian} are estimated by
\begin{equation}
\hat{\mu}
\;=\;
\frac{1}{T}\sum_{t=1}^T R_t,
\qquad
\hat{\sigma}^2
\;=\;
\frac{1}{T}\sum_{t=1}^T (R_t-\hat{\mu})^2.
\label{eq:mle_mu_sigma}
\end{equation}
The variance estimator in \eqref{eq:mle_mu_sigma} corresponds to the maximum likelihood estimator under the Gaussian assumption. This choice is consistent with the simulation-based estimators used throughout the paper, which rely on repeated sampling from the fitted nominal distribution. In particular, the use of the $1/T$ normalization (rather than $1/(T-1)$) ensures that $\hat{\sigma}^2$ is the likelihood-consistent variance parameter for the normal model.

\subsection{Role of the Nominal Model}
The estimated parameters $(\hat{\mu},\hat{\sigma})$ define the nominal return distribution
\begin{equation}
R \sim \mathcal{N}(\hat{\mu},\hat{\sigma}^2)
\quad \text{under } \mathbb{P},
\label{eq:fitted_nominal}
\end{equation}
which plays a central role in both VaR estimation methods considered in this study:
\begin{itemize}
\item In the \emph{importance sampling} approach, \eqref{eq:fitted_nominal} serves as the reference distribution with respect to which likelihood ratios are computed. A tilted sampling distribution is introduced to oversample tail losses, but unbiasedness of the estimator is defined relative to $\mathbb{P}$.
\item In the \emph{discrete moment matching} approach, the first four moments implied by \eqref{eq:fitted_nominal} (mean, variance, skewness, and kurtosis) are matched by a discrete distribution, which is then used to approximate the loss distribution for Monte Carlo VaR estimation.
\end{itemize}
In both cases, the Gaussian nominal model is treated as the \emph{assumed} data-generating process for the estimator, even though the true data-generating process used in the simulation experiments differs from \eqref{eq:fitted_nominal}.

\subsection{Closed-Form Gaussian VaR as a Pilot Estimate}
Under the nominal model \eqref{eq:fitted_nominal}, the loss variable $L=-R$ is also Gaussian with mean $-\hat{\mu}$ and variance $\hat{\sigma}^2$. The corresponding VaR at confidence level $\alpha$ admits the closed-form expression
\begin{equation}
x_0
\;=\;
\mathrm{VaR}_\alpha^{(\mathrm{Gauss})}
\;=\;
-\big(\hat{\mu} + \hat{\sigma} z_{1-\alpha}\big),
\label{eq:gaussian_var_closed}
\end{equation}
where $z_{1-\alpha}$ denotes the $(1-\alpha)$-quantile of the standard normal distribution. This quantity is used as a \emph{pilot VaR estimate} to guide the construction of the importance sampling change of measure. Importantly, \eqref{eq:gaussian_var_closed} is not treated as a final risk estimate; rather, it serves as an analytically convenient reference point that localizes the rare-event region $\{L>x\}$ for subsequent simulation.

\subsection{Assumptions and Limitations}
The nominal Gaussian model \eqref{eq:nominal_gaussian} rests on several simplifying assumptions:
\begin{enumerate}
\item The return sequence is assumed to be independent and identically distributed.
\item The distribution of returns is assumed to be light-tailed with finite exponential moments.
\item Parameters $(\mu,\sigma)$ are assumed to be time-invariant over the calibration window.
\end{enumerate}
These assumptions are deliberately relaxed in the simulation design by sampling returns from a heavy-tailed Student-$t$ distribution while retaining \eqref{eq:nominal_gaussian} as the estimator’s reference model. This deliberate mismatch allows the robustness of the VaR estimators to tail misspecification to be studied explicitly.

\section{True Data-Generating Process}
\label{sec:dgp}

\subsection{Distinction Between Nominal and True Measures}
A central feature of this study is the deliberate separation between the \emph{nominal} probability measure used by the VaR estimators and the \emph{true} probability measure governing the data-generating process. While both importance sampling and discrete moment matching are constructed under the Gaussian nominal model defined in Section~\ref{sec:calibration}, performance is evaluated under a heavy-tailed return distribution. This design allows the robustness of each estimator to tail misspecification to be assessed in a controlled manner.

To formalize this distinction, let $\mathbb{P}$ denote the nominal Gaussian measure under which estimators are calibrated, and let $\mathbb{P}^\star$ denote the true probability measure under which returns are generated in the simulation experiments.

\subsection{Student-$t$ Return Model}
Under the true measure $\mathbb{P}^\star$, one-day returns are assumed to follow a standardized Student-$t$ distribution:
\begin{equation}
R^\star \sim t_\nu(\mu^\star, s_\nu^2),
\label{eq:student_t_model}
\end{equation}
where $\nu>2$ denotes the degrees of freedom, $\mu^\star$ is the location parameter, and $s_\nu^2$ is the scale parameter. The Student-$t$ distribution is symmetric and heavy-tailed, with polynomial tail decay governed by $\nu$. As $\nu \to \infty$, the distribution converges to a Gaussian, whereas smaller values of $\nu$ correspond to increasingly heavy tails.

In all experiments, the location parameter is set equal to the nominal mean estimate,
\begin{equation}
\mu^\star := \hat{\mu},
\end{equation}
ensuring that misspecification arises solely from differences in tail behavior rather than shifts in central tendency.

\subsection{Variance Matching}
To isolate the effect of tail thickness, the Student-$t$ distribution is scaled to match the variance of the nominal Gaussian model. For $\nu>2$, the variance of a Student-$t$ random variable with scale parameter $s_\nu^2$ is given by
\begin{equation}
\mathrm{Var}_{\mathbb{P}^\star}(R^\star)
\;=\;
s_\nu^2 \,\frac{\nu}{\nu-2}.
\end{equation}
The scale parameter is therefore chosen such that
\begin{equation}
s_\nu^2
\;=\;
\hat{\sigma}^2 \,\frac{\nu-2}{\nu},
\label{eq:variance_matching}
\end{equation}
which ensures
\begin{equation}
\mathrm{Var}_{\mathbb{P}^\star}(R^\star)
\;=\;
\hat{\sigma}^2.
\end{equation}
This variance-matching step guarantees that any differences in VaR estimation accuracy across methods are attributable to higher-order tail behavior rather than differences in volatility scaling.

\subsection{Loss Variable Under the True Measure}
As in the nominal setting, the loss variable under the true measure is defined by
\begin{equation}
L^\star := -R^\star.
\label{eq:true_loss_def}
\end{equation}
The target risk quantity in the simulation study is the true VaR,
\begin{equation}
x_\alpha^\star
\;:=\;
\inf\big\{x : \mathbb{P}^\star(L^\star \le x) \ge \alpha \big\},
\label{eq:true_var_def}
\end{equation}
which generally differs from the Gaussian VaR implied by the nominal model in Section~\ref{sec:calibration}. In particular, for finite $\nu$, the heavy-tailed nature of the Student-$t$ distribution implies
\begin{equation}
x_\alpha^\star \;>\; x_\alpha^{(\mathrm{Gauss})}
\qquad \text{for large } \alpha,
\end{equation}
reflecting increased tail risk relative to the Gaussian benchmark.

\subsection{Experimental Regimes}
The simulation study considers a range of degrees of freedom
\begin{equation}
\nu \in \{5,\,7,\,10\},
\end{equation}
spanning moderately heavy-tailed to near-Gaussian regimes. For each value of $\nu$, independent samples are generated from \eqref{eq:student_t_model}, and VaR is estimated using methods calibrated under the nominal Gaussian model. This design enables a systematic examination of how estimator bias, variance, and stability evolve as tail thickness increases.

\subsection{Interpretation as Controlled Misspecification}
It is emphasized that the Student-$t$ model \eqref{eq:student_t_model} is not proposed as a realistic description of market returns. Rather, it serves as a mathematically tractable mechanism for introducing controlled deviations from Gaussian tail behavior. By holding the mean and variance fixed while varying $\nu$, the simulation framework isolates the impact of tail misspecification on VaR estimation, allowing the efficiency-robustness trade-off between importance sampling and discrete moment matching to be evaluated transparently.

\section{Value-at-Risk Estimation Methods}
\label{sec:methods}

This section defines the two VaR estimators studied in this paper. Throughout, $\alpha\in(0,1)$ is fixed (e.g.\ $\alpha=0.99$). The return random variable is denoted by $R$, and the associated loss is $L=-R$ as in \eqref{eq:loss_def}. The defining VaR equation is the tail-probability root condition
\begin{equation}
p(x_\alpha) = 1-\alpha,
\qquad
p(x) := \mathbb{P}(L>x)=\mathbb{P}(R<-x),
\label{eq:var_root_methods}
\end{equation}
where $\mathbb{P}$ denotes the nominal measure defined in Section~\ref{sec:calibration}.

\subsection{Discrete Moment Matching (DMM) via Moment-Feasible VaR Bracketing}
\label{subsec:dmm}

\subsubsection{Moment Information}
Let $L$ denote the (unknown) one-day loss random variable under the nominal measure $\mathbb{P}$. Suppose that the first $d$ raw moments of $L$ are available:
\begin{equation}
\mu_r := \mathbb{E}_{\mathbb{P}}[L^r], 
\qquad r=1,2,\dots,d,
\label{eq:raw_moments}
\end{equation}
with the convention $\mu_0 := 1$. In implementation, the moment vector $(\mu_1,\ldots,\mu_d)$ is computed from
loss samples generated under the \emph{nominal Gaussian model} calibrated in
Section~3. The DMM procedure therefore operates under the same nominal
information set as the importance sampling estimator. When the true
data-generating process differs from the nominal model, the imposed moment
constraints may themselves be misspecified; in this case, the resulting VaR
bounds should be interpreted as robust to tail uncertainty conditional on
nominal moment information.

\subsubsection{Discretization Grid and Moment-Feasible Set}
Fix an ordered finite grid (support) of possible loss values
\begin{equation}
X_0 < X_1 < \cdots < X_m,
\qquad
X_i \in \mathbb{R},
\label{eq:loss_grid}
\end{equation}
and consider discrete probability vectors
\begin{equation}
p = (p_0,\dots,p_m) \in \mathbb{R}^{m+1},
\qquad
p_i \ge 0,\ \ \sum_{i=0}^m p_i = 1.
\label{eq:prob_vector}
\end{equation}
The \emph{moment-feasible set} is defined as
\begin{equation}
\mathcal{P}_d
:=
\left\{
p \in \mathbb{R}^{m+1} :
p_i \ge 0,\ 
\sum_{i=0}^m p_i = 1,\ 
\sum_{i=0}^m X_i^r\, p_i = \mu_r,\ r=1,\dots,d
\right\}.
\label{eq:moment_feasible_set}
\end{equation}
Any $p\in\mathcal{P}_d$ defines a discrete distribution supported on the grid $\{X_i\}_{i=0}^m$ that matches the first $d$ raw moments of $L$ exactly.

\subsubsection{Quantile Functional on the Grid}
For a given feasible vector $p\in\mathcal{P}_d$, define its discrete cumulative distribution function (CDF) by
\begin{equation}
F_p(x) := \sum_{i: X_i \le x} p_i,
\label{eq:discrete_cdf}
\end{equation}
and define the $\alpha$-quantile (VaR functional) as
\begin{equation}
h_\alpha(p) 
:= 
\mathrm{VaR}_\alpha(p)
:=
\inf\{x\in\mathbb{R}: F_p(x)\ge \alpha\}.
\label{eq:quantile_functional}
\end{equation}
Because the support is finite and ordered, \eqref{eq:quantile_functional} reduces to
\begin{equation}
\mathrm{VaR}_\alpha(p)=X_{j^\star(p)},
\qquad
j^\star(p):=\min\left\{j\in\{0,\dots,m\}:\ \sum_{i=0}^j p_i \ge \alpha\right\}.
\label{eq:quantile_index}
\end{equation}
The map $p\mapsto \mathrm{VaR}_\alpha(p)$ is nonlinear due to the thresholding implicit in the cumulative sums.

\subsubsection{DMM Output as a VaR Interval (Deterministic Bracketing)}
The DMM philosophy adopted here is \emph{distributional ambiguity given finitely many moments}. Since the true distribution of $L$ is unknown beyond the moment constraints, the quantile itself can vary across $\mathcal{P}_d$. The tight moment-consistent VaR lower and upper bounds are defined by
\begin{equation}
v^-_\alpha := \inf_{p\in\mathcal{P}_d} \mathrm{VaR}_\alpha(p),
\qquad
v^+_\alpha := \sup_{p\in\mathcal{P}_d} \mathrm{VaR}_\alpha(p).
\label{eq:var_bounds_def}
\end{equation}
Any loss distribution consistent with \eqref{eq:moment_feasible_set} must satisfy
\begin{equation}
v^-_\alpha \le \mathrm{VaR}_\alpha(L) \le v^+_\alpha.
\label{eq:var_bracket}
\end{equation}
Thus, the DMM method returns an \emph{interval-valued} characterization of VaR, quantifying the range of tail risk compatible with the moment information.

\subsubsection{Computational Reduction via CDF Bounds}
Directly optimizing \eqref{eq:var_bounds_def} is difficult because the quantile functional is nonlinear. A standard reduction is to work with CDF bounds at fixed thresholds. For any fixed $x\in\mathbb{R}$, define the linear functional
\begin{equation}
F_p(x) = \sum_{i: X_i \le x} p_i,
\label{eq:cdf_linear}
\end{equation}
and define the moment-consistent lower and upper CDF envelopes
\begin{equation}
F^-(x) := \inf_{p\in\mathcal{P}_d} F_p(x),
\qquad
F^+(x) := \sup_{p\in\mathcal{P}_d} F_p(x).
\label{eq:cdf_bounds}
\end{equation}
Because \eqref{eq:cdf_linear} is linear in $p$ and $\mathcal{P}_d$ is defined by linear equalities/inequalities, both optimization problems in \eqref{eq:cdf_bounds} are linear programs.

The VaR bounds are then obtained by inversion:
\begin{equation}
v^-_\alpha = \inf\{x:\ F^-(x)\ge \alpha\},
\qquad
v^+_\alpha = \inf\{x:\ F^+(x)\ge \alpha\}.
\label{eq:invert_cdf_bounds}
\end{equation}
In implementation, $x$ is restricted to the grid $\{X_0,\dots,X_m\}$ and the inversion is computed by identifying the first grid point at which the bound crosses $\alpha$.

\subsubsection{Assumptions and Limitations of DMM}
The DMM bracketing procedure relies on:
\begin{enumerate}
\item \textbf{Finite-moment information:} only $\{\mu_r\}_{r=1}^d$ is assumed known. This choice is deliberate and defines the ambiguity set \eqref{eq:moment_feasible_set}.
\item \textbf{Grid restriction:} feasibility and bounds are computed only over distributions supported on the fixed grid \eqref{eq:loss_grid}. Therefore, the resulting interval \eqref{eq:var_bracket} is conditional on the discretization.
\item \textbf{Moment feasibility:} for large $d$ the constraint system may become infeasible
due to a combination of numerical ill-conditioning and theoretical moment
nonexistence under heavy-tailed distributions. In particular, when the true
data-generating process exhibits polynomial tails, higher-order moments
implied by the nominal model may not exist under the true measure.
\end{enumerate}
The method is distribution-free beyond the moment constraints, but it is not model-free in the sense that results depend on the chosen grid and the order of moments enforced.

\subsection{Importance Sampling (IS) via Exponential Tilting and VaR Root-Finding}
\label{subsec:is}

\subsubsection{Nominal Model and Tail Probability Target}
Under the nominal measure $\mathbb{P}$, returns satisfy
\begin{equation}
R \sim \mathcal{N}(\mu,\sigma^2),
\label{eq:is_nominal}
\end{equation}
with $(\mu,\sigma)$ estimated as in \eqref{eq:mle_mu_sigma}. The tail probability corresponding to a loss threshold $x$ is
\begin{equation}
p(x) 
:= \mathbb{P}(L>x)
= \mathbb{P}(R<-x)
= \mathbb{E}_{\mathbb{P}}\!\left[\mathbf{1}\{R<-x\}\right].
\label{eq:is_tail_prob_def}
\end{equation}
VaR is defined as the solution $x_\alpha$ to $p(x_\alpha)=1-\alpha$.

\subsubsection{Importance Sampling Measure}
To reduce variance when $1-\alpha$ is small, importance sampling introduces an alternative sampling measure $\mathbb{Q}_\theta$ under which the rare event $\{R<-x\}$ occurs more frequently. In this paper, $\mathbb{Q}_\theta$ is chosen as a mean-shifted Gaussian with unchanged variance:
\begin{equation}
R \sim \mathcal{N}(\mu-\theta,\sigma^2)
\quad \text{under } \mathbb{Q}_\theta,
\qquad \theta>0.
\label{eq:is_proposal}
\end{equation}
This is an exponential tilting (Esscher-type) change of measure for the Gaussian family.

\subsubsection{Likelihood Ratio (Radon-Nikodym Derivative)}
Let $f_{\mathbb{P}}$ and $f_{\mathbb{Q}_\theta}$ denote the densities of $R$ under \eqref{eq:is_nominal} and \eqref{eq:is_proposal}, respectively. The likelihood ratio (Radon-Nikodym derivative) is
\begin{equation}
W_\theta(r)
:= \frac{f_{\mathbb{P}}(r)}{f_{\mathbb{Q}_\theta}(r)}.
\label{eq:lr_def}
\end{equation}
For Gaussian densities with identical variance and a mean shift $\theta$, the ratio is available in closed form:
\begin{equation}
W_\theta(r)
=
\exp\!\left(
\frac{\theta}{\sigma^2}(r-\mu) + \frac{\theta^2}{2\sigma^2}
\right).
\label{eq:lr_closed}
\end{equation}
This identity is exact and is the mathematical justification for the weight computation in the code.

\subsubsection{Unbiased IS Estimator of the Tail Probability}
By change-of-measure,
\begin{equation}
p(x)
=
\mathbb{E}_{\mathbb{P}}[\mathbf{1}\{R<-x\}]
=
\mathbb{E}_{\mathbb{Q}_\theta}\!\left[\mathbf{1}\{R<-x\}W_\theta(R)\right].
\label{eq:com_identity}
\end{equation}
Given i.i.d.\ samples $R^{(1)},\dots,R^{(N)}\sim \mathcal{N}(\mu-\theta,\sigma^2)$ under $\mathbb{Q}_\theta$, define
\begin{equation}
\widehat{p}_{\mathrm{IS}}(x;\theta)
:=
\frac{1}{N}\sum_{i=1}^N \mathbf{1}\{R^{(i)}<-x\}\,W_\theta(R^{(i)}).
\label{eq:is_estimator}
\end{equation}
Then $\mathbb{E}_{\mathbb{Q}_\theta}[\widehat{p}_{\mathrm{IS}}(x;\theta)] = p(x)$, i.e.\ \eqref{eq:is_estimator} is unbiased for the nominal tail probability $p(x)$.

\subsubsection{Tilt Selection (Glasserman-Style Heuristic)}
The efficiency of \eqref{eq:is_estimator} depends on choosing $\theta$ so that the proposal distribution places non-negligible mass near the rare-event boundary. A practical heuristic is to use the closed-form Gaussian pilot VaR
\begin{equation}
x_0 = -(\mu+\sigma z_{1-\alpha}),
\label{eq:is_pilot}
\end{equation}
and choose $\theta$ to shift the mean under $\mathbb{Q}_\theta$ toward the boundary return level $-x_0$:
\begin{equation}
\mu-\theta \approx -x_0
\quad \Longrightarrow \quad
\theta := \mu + x_0.
\label{eq:is_theta_choice}
\end{equation}
This choice increases the frequency of samples satisfying $R<-x$ when $x$ is near the VaR region, thereby reducing the variance of \eqref{eq:is_estimator} relative to naive Monte Carlo.

\subsubsection{VaR Computation via Root-Finding (Bracketing + Bisection)}
Since $p(x)$ is monotone decreasing in $x$, the VaR equation can be solved numerically. Define the target tail probability
\begin{equation}
\pi_\alpha := 1-\alpha.
\label{eq:target_tail}
\end{equation}
The IS-based VaR estimate is defined as the numerical solution to
\begin{equation}
\widehat{p}_{\mathrm{IS}}(x;\theta) = \pi_\alpha.
\label{eq:is_root_equation}
\end{equation}
A bracketing interval $[x_{\min},x_{\max}]$ is first constructed such that
\begin{equation}
\widehat{p}_{\mathrm{IS}}(x_{\min};\theta)\ge \pi_\alpha,
\qquad
\widehat{p}_{\mathrm{IS}}(x_{\max};\theta)\le \pi_\alpha.
\label{eq:bracket_condition}
\end{equation}
Given a valid bracket, bisection iteratively updates $x_{\min},x_{\max}$ until $x_{\max}-x_{\min}$ is below a tolerance. The IS VaR estimate returned by the procedure is
\begin{equation}
\widehat{\mathrm{VaR}}_{\alpha,\mathrm{IS}} := \frac{x_{\min}+x_{\max}}{2}.
\label{eq:is_var_output}
\end{equation}

\subsubsection{Common Random Numbers (CRN) for Numerical Stability}
Because $\widehat{p}_{\mathrm{IS}}(x;\theta)$ is a Monte Carlo estimator, it is noisy as a function of $x$. To prevent bisection from reacting to independent simulation noise at each iteration, common random numbers are used: the random number generator is reset to the same seed at each bisection evaluation, so that differences in $\widehat{p}_{\mathrm{IS}}(x;\theta)$ across nearby $x$ values reflect changes in $x$ rather than unrelated Monte Carlo variation. This stabilizes the monotone behavior required by \eqref{eq:bracket_condition}.

\subsubsection{Weight Stability Diagnostics}
Importance sampling can fail numerically when weights become highly concentrated (weight degeneracy), leading to inflated variance and unreliable estimates. Two diagnostics are recorded.

First define normalized weights
\begin{equation}
\widetilde{w}_i
:=
\frac{W_\theta(R^{(i)})}{\sum_{j=1}^N W_\theta(R^{(j)})},
\qquad i=1,\dots,N.
\label{eq:normalized_weights}
\end{equation}
The effective sample size (ESS) proxy is
\begin{equation}
\mathrm{ESS}
:=
\frac{1}{\sum_{i=1}^N \widetilde{w}_i^2},
\label{eq:ess_def}
\end{equation}
which lies in $[1,N]$ and measures weight dispersion (larger is better). The maximum normalized weight share is
\begin{equation}
w_{\max} := \max_{1\le i\le N} \widetilde{w}_i,
\label{eq:max_share}
\end{equation}
which quantifies whether the estimator is dominated by a small number of extreme samples. Large values of $w_{\max}$ and very small values of ESS indicate potential degeneracy.

\subsubsection{Assumptions and Limitations of IS}
The IS estimator \eqref{eq:is_estimator} is unbiased for $p(x)$ under the nominal measure $\mathbb{P}$ provided that:
\begin{enumerate}
\item The proposal density $f_{\mathbb{Q}_\theta}$ is absolutely continuous with respect to $f_{\mathbb{P}}$ (true for Gaussian mean shifts).
\item The likelihood ratio \eqref{eq:lr_closed} is evaluated exactly and without overflow/underflow.
\item The second moment $\mathbb{E}_{\mathbb{Q}_\theta}\!\big[\mathbf{1}\{R<-x\}W_\theta(R)^2\big]$ is finite (which holds in the Gaussian setting but can still yield severe finite-sample degeneracy for aggressive tilts).
\end{enumerate}
In the misspecification experiments of Section~\ref{sec:dgp}, performance is evaluated under $\mathbb{P}^\star$ (Student-$t$ returns). In that setting, the IS procedure remains a valid estimator of the \emph{nominal-model} VaR but may exhibit bias relative to the \emph{true} VaR $x_\alpha^\star$ because the Gaussian nominal model does not match the true tail behavior. This bias is an intended object of study and is measured explicitly in the results section.

\section{Simulation Design}
\label{sec:simulation}

This section describes the simulation framework used to evaluate the performance of the VaR estimators defined in Section~\ref{sec:methods}. The design is constructed to ensure a fair and controlled comparison between importance sampling and discrete moment matching under varying degrees of tail misspecification.

\subsection{Experimental Parameters}
The simulation study considers the following fixed parameters:
\begin{itemize}
\item \textbf{Confidence level:} $\alpha \in \{0.99\}$.
\item \textbf{Degrees of freedom:} $\nu \in \{5,7,10\}$ for the Student-$t$ data-generating process defined in Section~\ref{sec:dgp}.
\item \textbf{Monte Carlo budget:} $N$ samples per VaR evaluation (e.g.\ $N=10{,}000$).
\item \textbf{Number of replications:} $M$ independent simulation runs (e.g.\ $M=100$).
\end{itemize}
All reported performance metrics are computed across the $M$ replications.

\subsection{Simulation Loop}
For each degree of freedom $\nu \in \{5,7,10\}$ and each replication index $m=1,\dots,M$, the following steps are executed:

\begin{enumerate}
\item \textbf{Generate true returns.}  
Draw an i.i.d.\ sample
\begin{equation}
R^{\star,(m)}_1,\dots,R^{\star,(m)}_T \;\sim\; t_\nu(\hat{\mu}, s_\nu^2),
\label{eq:true_sample}
\end{equation}
where $s_\nu^2$ is defined by the variance-matching condition \eqref{eq:variance_matching}. This sample represents returns generated under the true measure $\mathbb{P}^\star$.

\item \textbf{Compute true losses.}  
Define
\begin{equation}
L^{\star,(m)}_t := -R^{\star,(m)}_t,
\qquad t=1,\dots,T.
\label{eq:true_losses}
\end{equation}

\item \textbf{Estimate nominal parameters.}  
Using the return sample $\{R^{\star,(m)}_t\}_{t=1}^T$, compute the Gaussian MLEs
\begin{equation}
\hat{\mu}^{(m)} := \frac{1}{T}\sum_{t=1}^T R^{\star,(m)}_t,
\qquad
(\hat{\sigma}^{(m)})^2 := \frac{1}{T}\sum_{t=1}^T (R^{\star,(m)}_t-\hat{\mu}^{(m)})^2.
\label{eq:reestimated_params}
\end{equation}
These parameters define the nominal model used by both VaR estimators in replication $m$.

\item \textbf{Compute true VaR benchmark.}  
The true VaR under $\mathbb{P}^\star$ is computed analytically using the Student-$t$ quantile:
\begin{equation}
x_{\alpha}^{\star,(m)}
:=
-\left(
\hat{\mu}^{(m)} + s_\nu \, t_{\nu,\,1-\alpha}
\right),
\label{eq:true_var_student}
\end{equation}
where $t_{\nu,\,1-\alpha}$ denotes the $(1-\alpha)$-quantile of a standardized Student-$t$ distribution with $\nu$ degrees of freedom.

\item \textbf{DMM VaR estimation.}  
Using the nominal moments implied by $\hat{\mu}^{(m)}$ and $\hat{\sigma}^{(m)}$, the DMM procedure described in Section~\ref{subsec:dmm} is applied to compute lower and upper VaR bounds
\begin{equation}
\widehat{v}^{-,(m)}_\alpha,
\qquad
\widehat{v}^{+,(m)}_\alpha.
\label{eq:dmm_output}
\end{equation}
When a point estimate is required for comparison, the midpoint
\begin{equation}
\widehat{x}^{(m)}_{\alpha,\mathrm{DMM}}
:=
\frac{\widehat{v}^{-,(m)}_\alpha + \widehat{v}^{+,(m)}_\alpha}{2}
\label{eq:dmm_point}
\end{equation}
is reported, with the interval width retained as a measure of ambiguity.

\item \textbf{IS VaR estimation.}  
Using the nominal Gaussian model \eqref{eq:is_nominal} with parameters $(\hat{\mu}^{(m)},\hat{\sigma}^{(m)})$, the IS procedure of Section~\ref{subsec:is} is applied:
\begin{enumerate}
\item Compute the pilot Gaussian VaR $x_0^{(m)}$ via \eqref{eq:is_pilot}.
\item Set the tilting parameter $\theta^{(m)}$ via \eqref{eq:is_theta_choice}.
\item Solve the root equation \eqref{eq:is_root_equation} by bracketing and bisection to obtain
\begin{equation}
\widehat{x}^{(m)}_{\alpha,\mathrm{IS}}.
\end{equation}
\item Record stability diagnostics: ESS and $w_{\max}$.
\end{enumerate}
\end{enumerate}

\subsection{Performance Metrics}
For each estimator and each value of $\nu$, performance is summarized using the following metrics computed across replications $m=1,\dots,M$:

\begin{itemize}
\item \textbf{Bias:}
\begin{equation}
\mathrm{Bias}
:=
\frac{1}{M}\sum_{m=1}^M
\left(
\widehat{x}^{(m)}_\alpha - x_{\alpha}^{\star,(m)}
\right).
\label{eq:bias_def}
\end{equation}

\item \textbf{Variance:}
\begin{equation}
\mathrm{Var}
:=
\frac{1}{M-1}\sum_{m=1}^M
\left(
\widehat{x}^{(m)}_\alpha - \overline{x}_\alpha
\right)^2,
\qquad
\overline{x}_\alpha := \frac{1}{M}\sum_{m=1}^M \widehat{x}^{(m)}_\alpha.
\label{eq:var_def_sim}
\end{equation}

\item \textbf{Mean squared error (MSE):}
\begin{equation}
\mathrm{MSE} := \mathrm{Bias}^2 + \mathrm{Var}.
\label{eq:mse_def}
\end{equation}

\item \textbf{IS stability diagnostics:}
\begin{equation}
\overline{\mathrm{ESS}} := \frac{1}{M}\sum_{m=1}^M \mathrm{ESS}^{(m)},
\qquad
\overline{w}_{\max} := \frac{1}{M}\sum_{m=1}^M w_{\max}^{(m)}.
\label{eq:is_diag_avg}
\end{equation}
\end{itemize}

For discrete moment matching, bias, variance, and mean squared error are
reported for the midpoint of the VaR interval solely as descriptive
summaries. These quantities are not interpreted as optimality criteria, as
DMM is designed to produce interval-valued guarantees that quantify
distributional ambiguity rather than point estimates.

\subsection{Reproducibility Considerations}
All random number generators are initialized with fixed seeds at the start of each replication to ensure reproducibility. Within the IS root-finding procedure, common random numbers are used across bisection iterations, as described in Section~\ref{subsec:is}, to stabilize the monotonic behavior of the estimated tail probability.

\subsection{Interpretation}
The simulation design ensures that both estimators operate under identical nominal assumptions and identical Monte Carlo budgets. Any observed differences in bias, variance, or stability therefore reflect structural differences between importance sampling and discrete moment matching, rather than artifacts of unequal calibration or simulation effort.

\section{Results}
\label{sec:results}

This section reports the empirical behavior of the VaR procedures defined in
Section~\ref{sec:methods} under the simulation design of
Section~\ref{sec:simulation}. The objective is not to compare two estimators of
the same quantity, but to study how two fundamentally different approaches to
tail risk, efficient estimation under a nominal model (importance sampling)
and robust bracketing under moment ambiguity (discrete moment matching), behave
as tail thickness increases.

All performance is evaluated under the heavy-tailed Student-$t$ data-generating
process described in Section~\ref{sec:dgp}, while the importance sampling (IS)
procedure remains calibrated to the nominal Gaussian model described in
Section~\ref{subsec:is}. Throughout, the loss is defined as $L=-R$, and the VaR
level $\alpha$ corresponds to the $(1-\alpha)$ upper tail quantile of $L$.

\subsection{Baseline Behavior Under the Nominal Gaussian Model}
\label{subsec:results_baseline}

As a preliminary validation, both IS and DMM were run under the nominal Gaussian model. In this setting, the closed-form Gaussian VaR provides an exact benchmark. The IS implementation reproduced the nominal Gaussian VaR closely, confirming the correctness of (i) the exponential tilting construction, (ii) likelihood ratio evaluation, and (iii) the bisection root-finding routine used to solve the VaR equation. Numerically, the bisection bracket remained valid throughout iterations, consistent with the monotonicity of the estimated tail probability in the loss threshold when common random numbers are used across bisection steps. This baseline check isolates subsequent performance differences under Student-$t$ returns as effects of tail misspecification rather than numerical implementation errors.

\subsection{Bias Under Heavy-Tailed Data-Generating Processes}
\label{subsec:results_bias}

When returns are generated under Student-$t$ tails, systematic differences
between the two approaches emerge due to their fundamentally different targets.
For fixed confidence level $\alpha$, decreasing the degrees of freedom $\nu$
increases the true VaR $x^\star_\alpha$, reflecting heavier tails.

Under these heavy-tailed scenarios, the IS procedure converges efficiently to
the VaR implied by the nominal Gaussian model, which lies below the true VaR
under $\mathbb{P}^\star$. As a result, the observed negative bias relative to
$x^\star_\alpha$ reflects structural model misspecification rather than
estimation error. Increasing the Monte Carlo budget reduces variance but cannot
eliminate this bias, as IS continues to target the nominal-model tail
probability rather than the true heavy-tailed tail probability.

Table~\ref{tab:is_summary} reports IS mean, standard deviation, and bias relative to the analytically computed Student-$t$ true VaR for $\nu\in\{5,7,10\}$ and $\alpha\in\{0.990,0.995\}$. The bias is negative in all settings and increases in magnitude as tails become heavier and as $\alpha$ increases.

\begin{table}[h!]
\centering
\caption{Importance sampling VaR performance under Student-$t$ true returns. Bias is defined as $\mathrm{Bias}=\mathrm{IS\_mean}-\mathrm{trueVaR}$. ESS denotes the effective sample size proxy computed from normalized weights, and maxW is the maximum normalized weight share.}
\label{tab:is_summary}
\begin{tabular}{c c c c c c c c}
\hline
$\nu$ & $\alpha$ & trueVaR & IS\_mean & IS\_std & IS\_bias & ESS & maxW \\
\hline
5  & 0.990 & 0.0281 & 0.0252 & 0.0017 & -0.0029 & 154 & 0.056 \\
5  & 0.995 & 0.0339 & 0.0283 & 0.0022 & -0.0056 &  91 & 0.081 \\
7  & 0.990 & 0.0273 & 0.0246 & 0.0013 & -0.0026 & 154 & 0.056 \\
7  & 0.995 & 0.0320 & 0.0276 & 0.0013 & -0.0044 &  91 & 0.081 \\
10 & 0.990 & 0.0266 & 0.0249 & 0.0010 & -0.0017 & 154 & 0.056 \\
10 & 0.995 & 0.0307 & 0.0276 & 0.0009 & -0.0030 &  91 & 0.081 \\
\hline
\end{tabular}
\end{table}

\subsection{Estimator Variability and Efficiency}
\label{subsec:results_variance}

Importance sampling yields relatively low estimator variability across replications. As shown in Table~\ref{tab:is_summary}, IS standard deviations remain below $2.5\times 10^{-3}$ even at $\alpha=0.995$, and decrease as $\nu$ increases (lighter tails). This behavior is consistent with the variance-reduction objective of IS: the change of measure increases the frequency of loss observations near the target tail region under the nominal model, thereby reducing Monte Carlo noise.

By contrast, DMM produces deterministic VaR \emph{intervals} rather than a single point estimate. The width of the interval reflects distributional ambiguity given the enforced moments and the discretization grid. Empirically, the DMM lower bound increases monotonically with the number of enforced raw moments $d$, while the upper bound remains nearly constant; hence the interval tightens and stabilizes as $d$ increases. This tightening pattern is visible in Figure~\ref{fig:dmm_sensitivity} and provides a numerical consistency check that additional moment constraints reduce ambiguity.

\subsection{Discrete Moment Matching: Bracketing, Sensitivity, and Feasibility}
\label{subsec:results_dmm}

\begin{figure}[h!]
\centering
\includegraphics[width=0.72\textwidth]{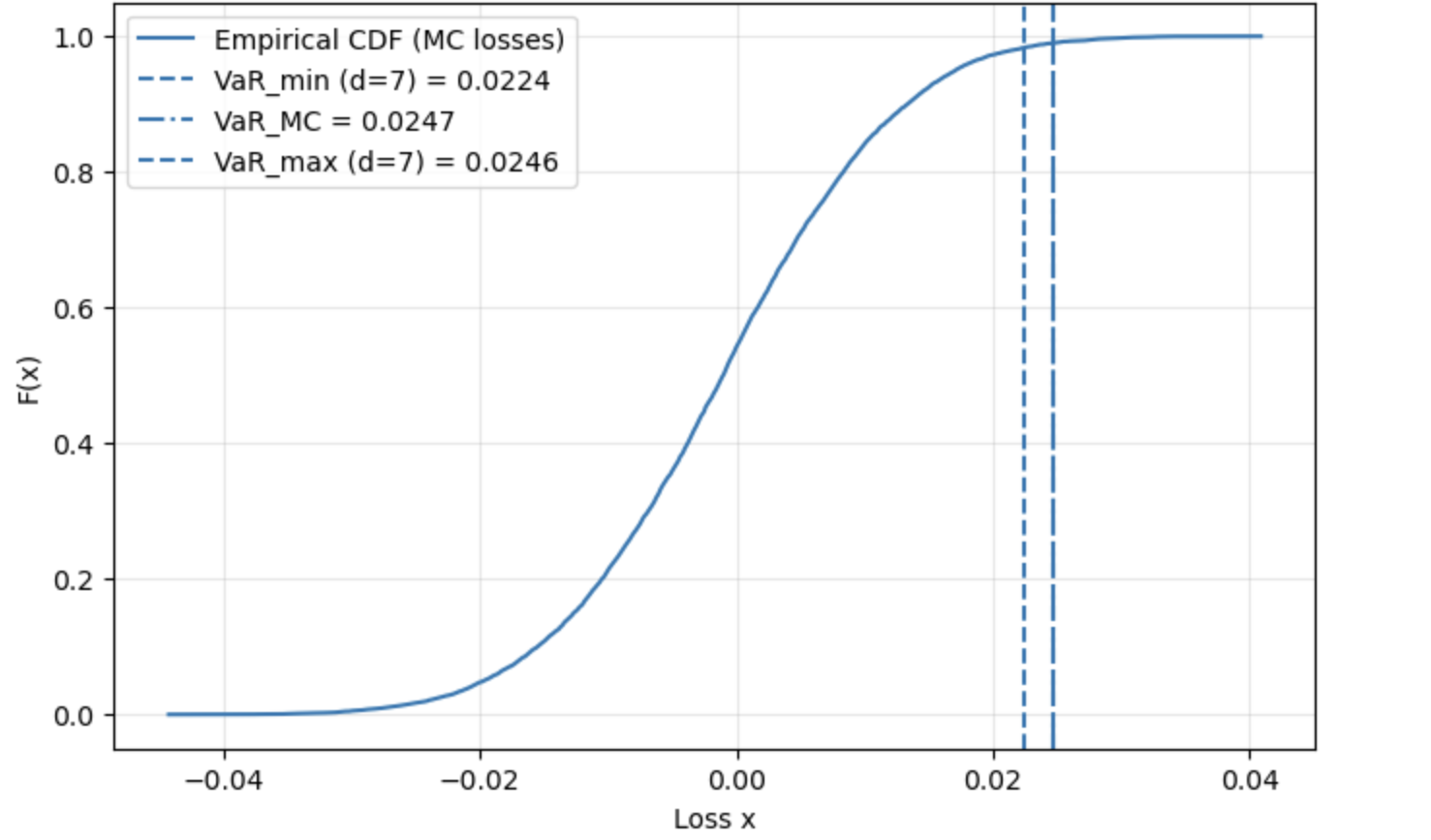}
\caption{Moment-feasible VaR bracketing versus empirical loss CDF at $\alpha=0.99$. The vertical lines indicate the DMM lower and upper VaR bounds at the chosen moment order and the Monte Carlo VaR reference.}
\label{fig:dmm_bracketing}
\end{figure}

Figure~\ref{fig:dmm_bracketing} overlays the empirical loss CDF with the DMM VaR bounds at $\alpha=0.99$. The vertical lines in the plot correspond to the DMM lower and upper VaR bounds computed at the selected moment order (reported in the code as $d^\star=7$) and a Monte Carlo VaR reference line. The figure provides a geometric interpretation of DMM: instead of fitting a parametric tail, the method computes quantile bounds that are compatible with the enforced moment constraints.

\begin{figure}[h!]
\centering
\includegraphics[width=0.65\textwidth]{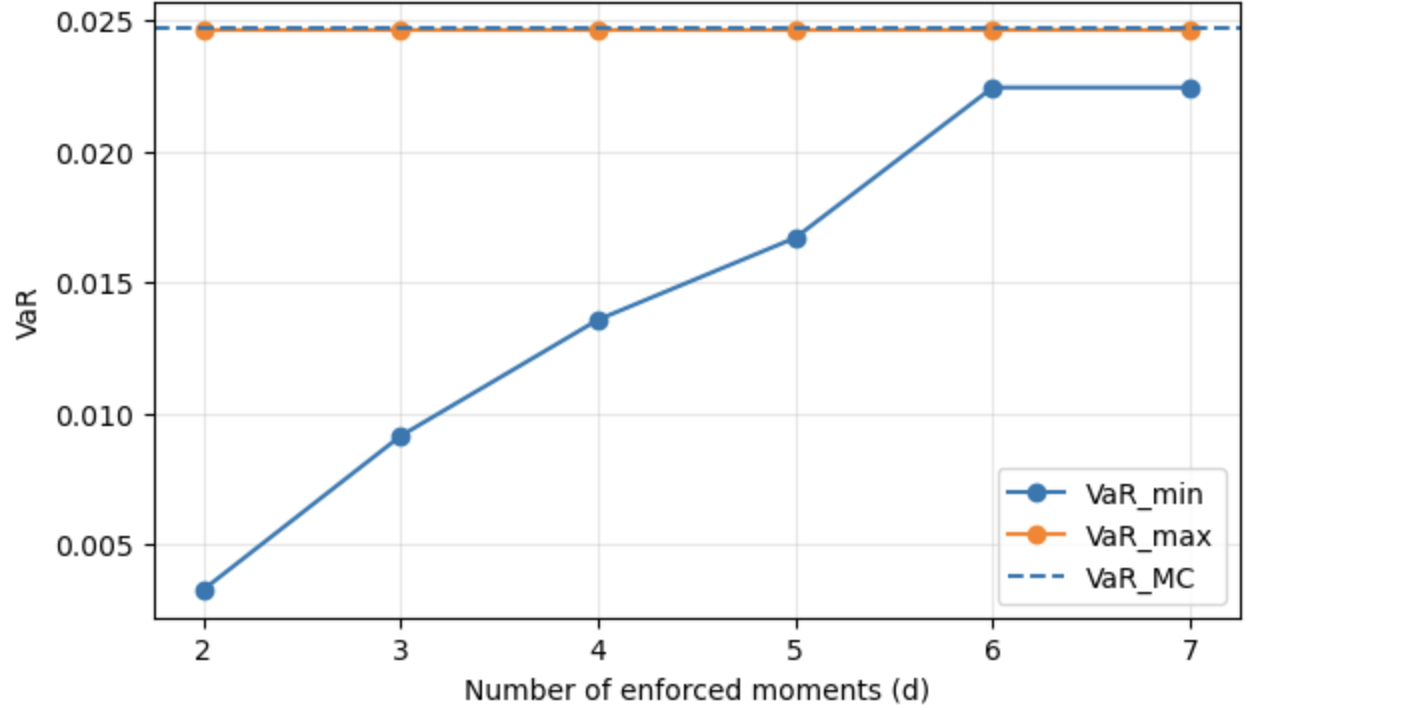}
\caption{Sensitivity of DMM VaR bounds to moment order $d$. The lower bound increases with $d$ and stabilizes, while the upper bound remains nearly constant.}
\label{fig:dmm_sensitivity}
\end{figure}

Figure~\ref{fig:dmm_sensitivity} shows sensitivity of the DMM VaR bounds to the number of enforced moments $d$. In the reported experiment, the upper bound remains essentially unchanged while the lower bound increases with $d$ and stabilizes for moderate values of $d$. This is consistent with contraction of the moment-feasible distribution set as additional constraints are imposed.

The code additionally reports a feasibility breakdown when enforcing higher
moment orders. Feasibility holds up to $d^\star=7$ and fails for $d \ge 8$ in the
reported experiments. This behavior should be interpreted carefully.
For Student-$t$ distributions with degrees of freedom $\nu$, raw moments of
order $r \ge \nu$ do not exist. In particular, when $\nu=5$, moments above the
fourth order are theoretically undefined.

Consequently, infeasibility at higher moment orders reflects a combination of
theoretical moment nonexistence and numerical ill-conditioning of the truncated
moment problem under discretization. The value $d^\star$ should therefore be
interpreted not as an optimal moment order, but as the largest order for which
the imposed moment constraints remain both theoretically meaningful and
numerically feasible within the chosen grid.

\subsection{Importance Sampling Stability Diagnostics}
\label{subsec:results_diagnostics}

Stability diagnostics provide insight into the alignment between the IS change
of measure and the geometry of the rare-event set under the true data-generating
process. In the reported experiments, the effective sample size (ESS) is
substantially smaller than the Monte Carlo budget, with ESS decreasing from
approximately $154$ at $\alpha=0.990$ to approximately $91$ at $\alpha=0.995$
(Table~\ref{tab:is_summary}). At the same time, the maximum normalized weight
share increases as $\alpha$ rises.

These patterns are characteristic of increasingly rare-event regimes and do
not, by themselves, imply numerical failure. Rather, they indicate that the
Gaussian exponential tilting measure becomes progressively less well aligned
with the true heavy-tailed loss distribution as confidence levels increase.

\begin{figure}[h!]
\centering
\includegraphics[width=0.65\textwidth]{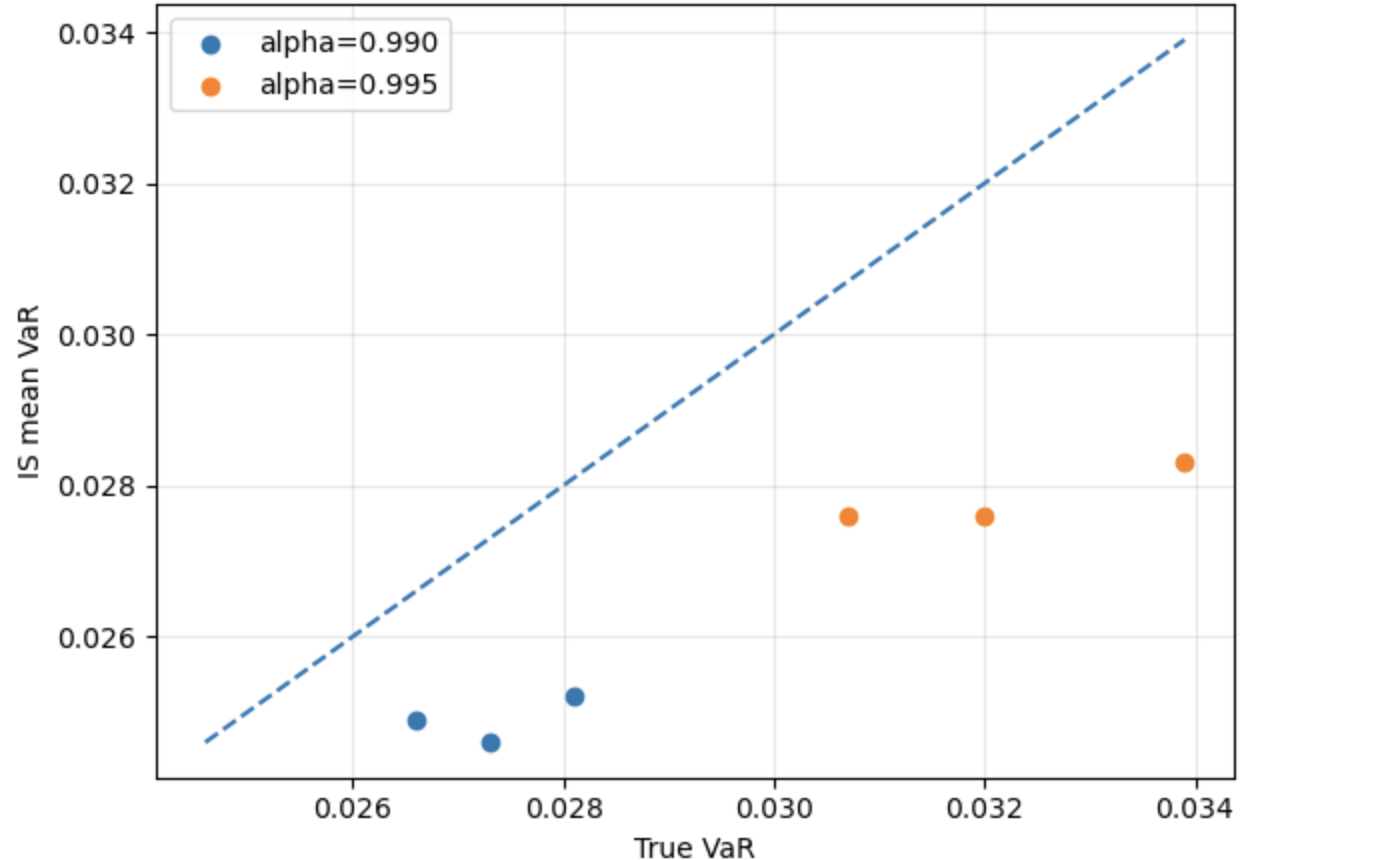}
\caption{Calibration of IS VaR under Student-$t$ tails: IS mean VaR versus true VaR. The dashed line is the identity line; points below indicate underestimation.}
\label{fig:is_calibration}
\end{figure}

\begin{figure}[h!]
\centering
\includegraphics[width=0.65\textwidth]{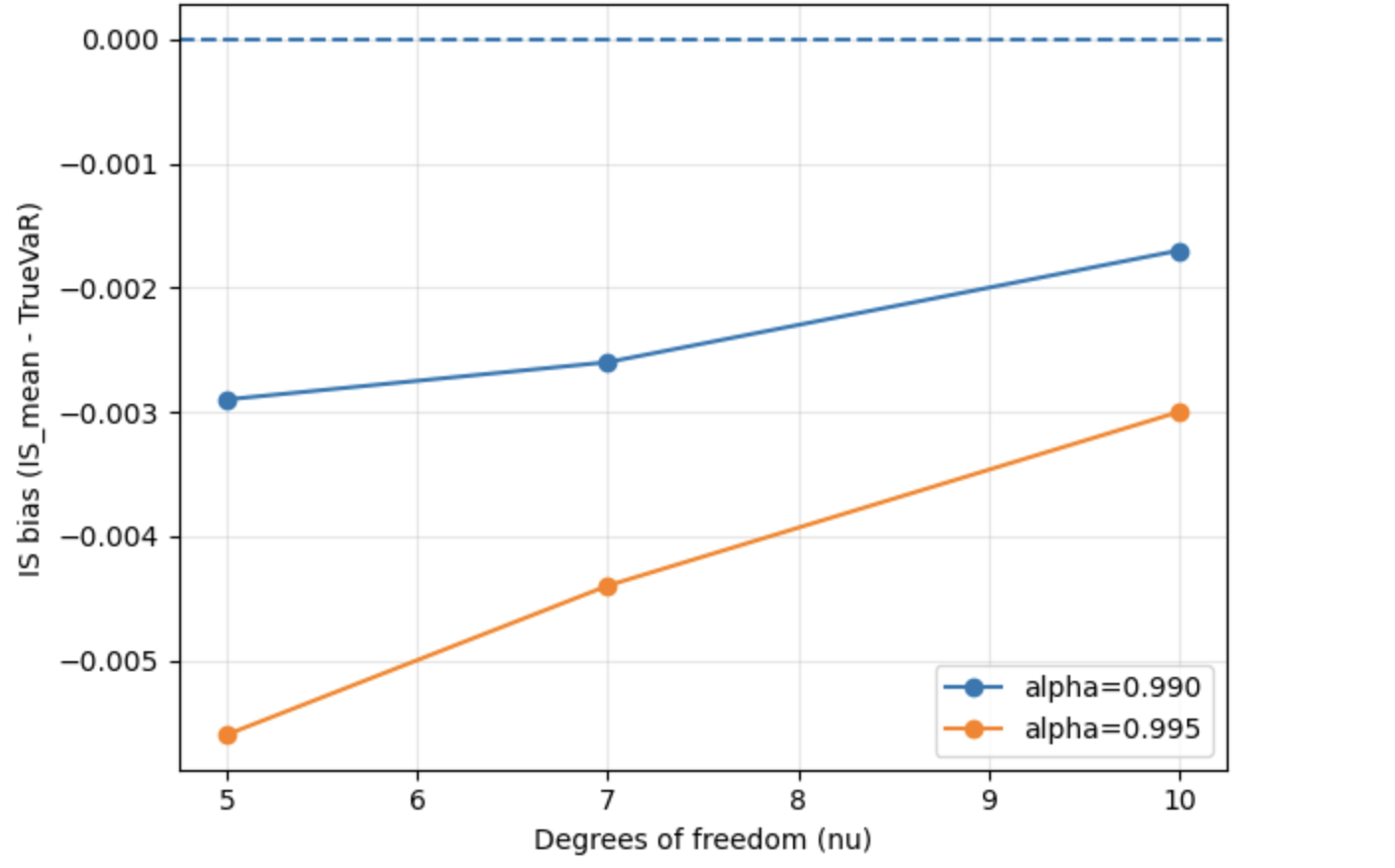}
\caption{IS bias (IS mean minus true VaR) versus degrees of freedom $\nu$. Bias magnitude decreases as tails become lighter.}
\label{fig:is_bias_nu}
\end{figure}

\begin{figure}[h!]
\centering
\includegraphics[width=0.65\textwidth]{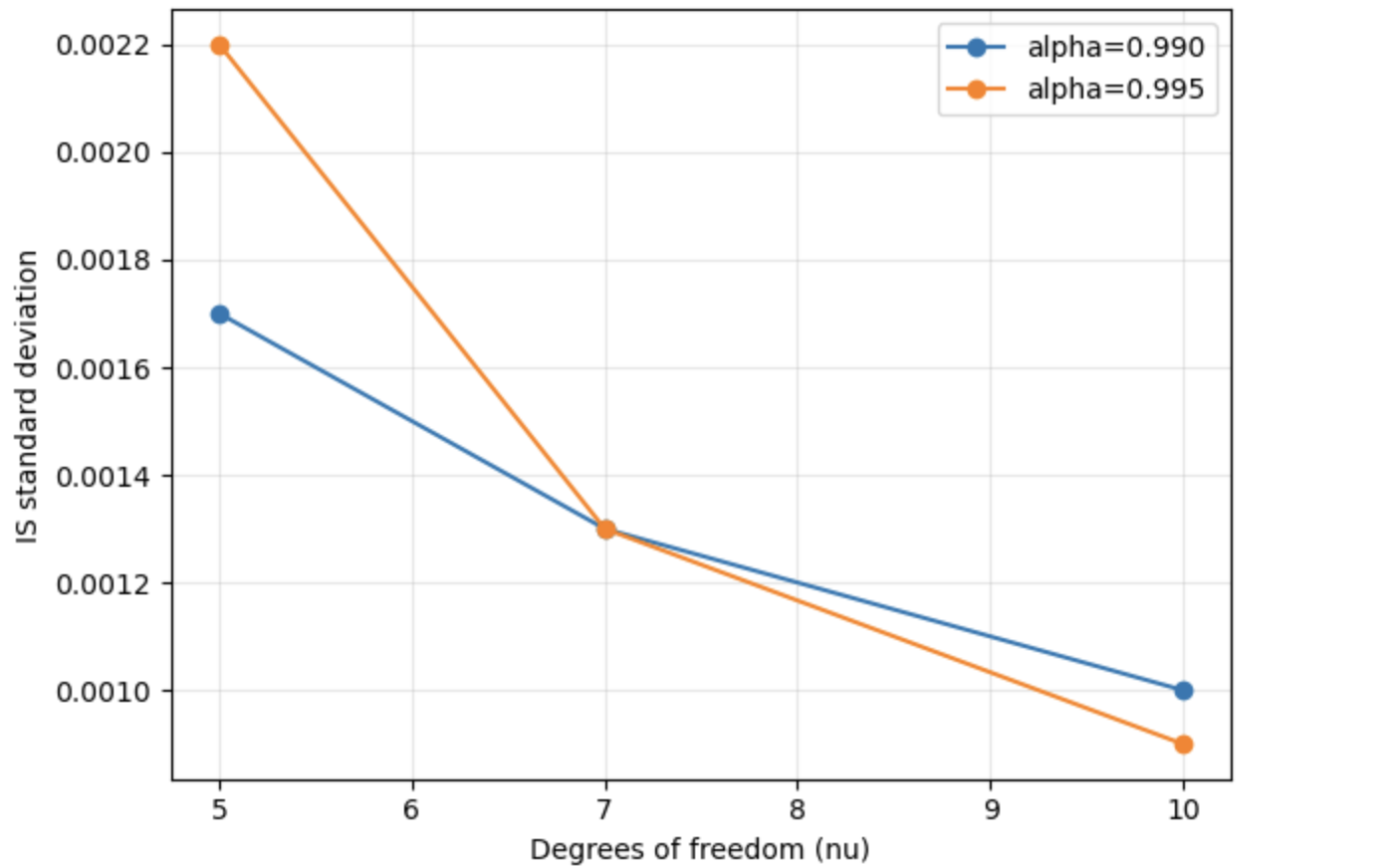}
\caption{IS estimator variability (standard deviation across replications) versus degrees of freedom $\nu$.}
\label{fig:is_std_nu}
\end{figure}

\begin{figure}[h!]
\centering
\includegraphics[width=0.75\textwidth]{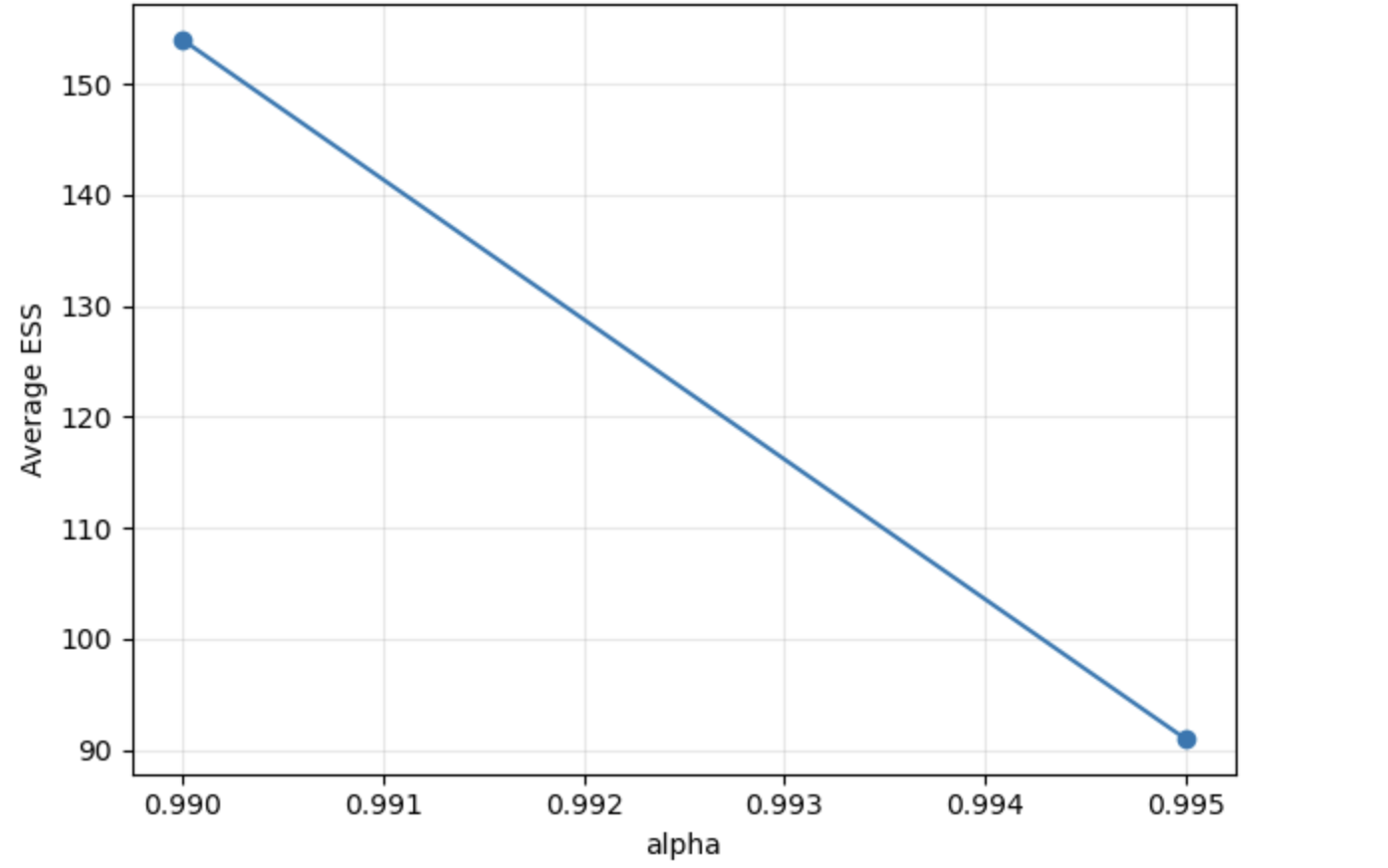}
\caption{IS stability diagnostics versus confidence level $\alpha$. Top: average ESS decreases from $\approx 154$ at $\alpha=0.990$ to $\approx 91$ at $\alpha=0.995$. Bottom: max weight share increases from $\approx 0.056$ to $\approx 0.081$, indicating stronger weight concentration at higher confidence levels.}
\label{fig:is_stability_alpha}
\end{figure}

\begin{figure}[h!]
\centering
\includegraphics[width=0.65\textwidth]{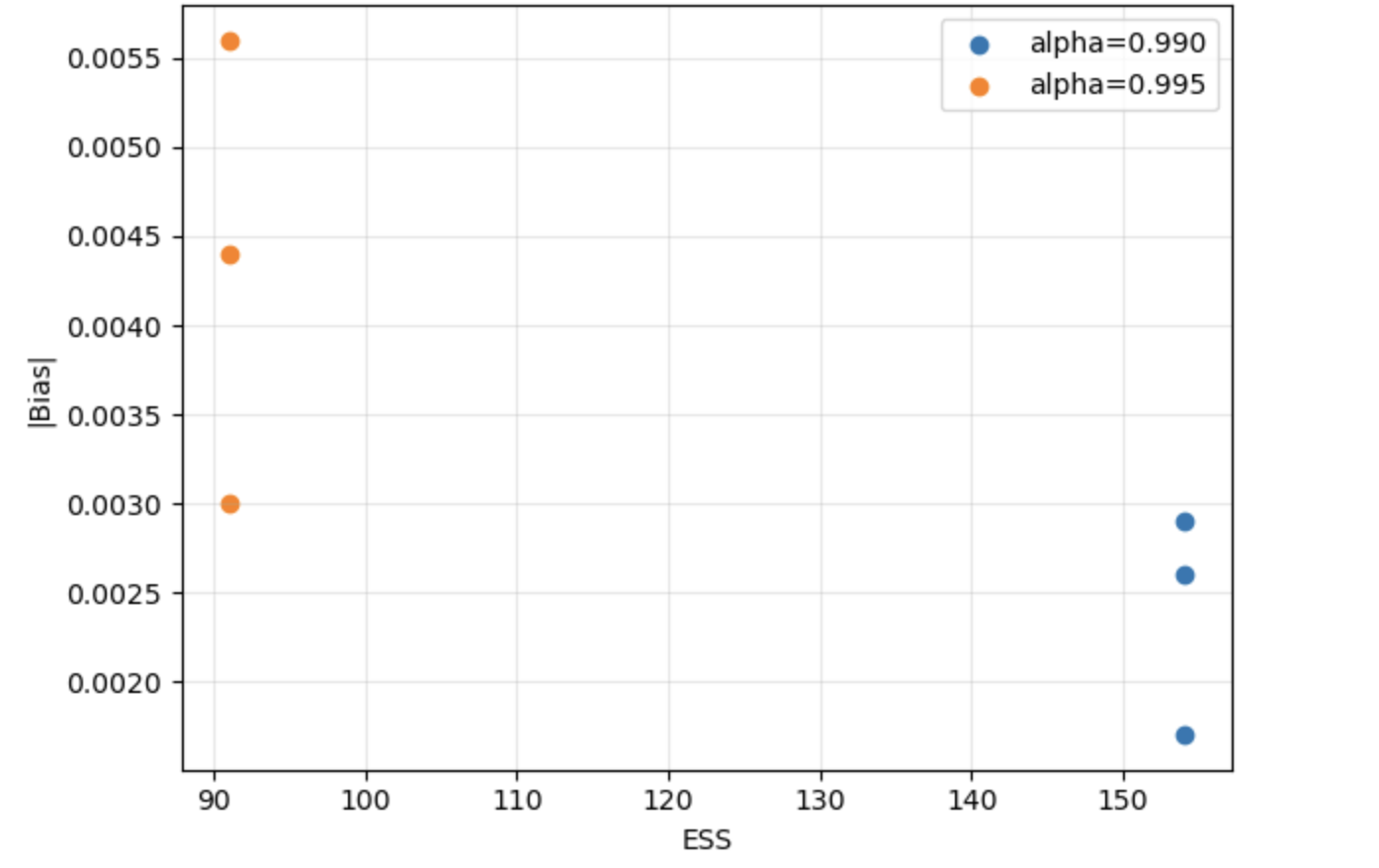}
\caption{Tradeoff diagnostic: absolute bias versus ESS. The higher-confidence regime exhibits lower ESS and larger absolute bias, reflecting increased rare-event difficulty and tail-model mismatch.}
\label{fig:is_tradeoff}
\end{figure}

The diagnostic plots produced by the code are reported in Figures~\ref{fig:is_calibration} to \ref{fig:is_tradeoff}. Figure~\ref{fig:is_calibration} shows that IS mean VaR values fall below the $45^\circ$ line when plotted against the true Student-$t$ VaR, visually confirming systematic underestimation under tail misspecification. Figure~\ref{fig:is_bias_nu} shows that the bias magnitude decreases as $\nu$ increases (tails become lighter), consistent with convergence toward Gaussian behavior. Figure~\ref{fig:is_stability_alpha} summarizes stability degradation as $\alpha$ increases, and Figure~\ref{fig:is_tradeoff} highlights that the higher-confidence regime ($\alpha=0.995$) is associated with both lower ESS and larger absolute bias.

\subsection{Summary of Findings}
\label{subsec:results_summary}

Taken together, the empirical results highlight a clear efficiency-robustness
trade-off. Importance sampling delivers low-variance estimates of VaR under the
nominal Gaussian model and exhibits stable numerical behavior across
replications. However, under Student-$t$ tails, it converges efficiently to the
nominal-model VaR, which systematically understates the true tail risk; the
resulting bias increases with heavier tails and higher confidence levels.

Discrete moment matching exhibits complementary behavior. Rather than producing
a point estimate, DMM yields deterministic VaR bounds that remain valid under
tail misspecification. As additional moment constraints are imposed, the
moment-feasible ambiguity set contracts and the VaR bounds tighten up to a
theoretically and numerically feasible limit. The resulting interval width
provides an explicit measure of distributional uncertainty, avoiding spurious
precision in settings where tail behavior is difficult to specify.

\section{Discussion}
\label{sec:discussion}

The results of Section~\ref{sec:results} highlight a fundamental trade-off in simulation-based Value-at-Risk estimation between computational efficiency and robustness to model misspecification. Although both importance sampling and discrete moment matching aim to improve upon naive Monte Carlo estimation in rare-event settings, they do so through fundamentally different mechanisms, and these differences become most apparent when tail behavior deviates from nominal modeling assumptions.

\subsection{Efficiency Versus Robustness}
Importance sampling is designed to reduce estimator variance by altering the sampling distribution so that rare events occur more frequently. When the nominal model accurately reflects the geometry of extreme losses, this approach can yield substantial efficiency gains. In the present study, importance sampling exhibits low variability and stable root-finding behavior under the Gaussian nominal model, confirming the correctness of the likelihood-ratio construction and numerical implementation.

However, the results demonstrate that variance reduction alone does not guarantee accurate risk estimation under tail misspecification. When the true return distribution exhibits heavier tails than the nominal Gaussian model, importance sampling remains efficient but converges to the \emph{wrong target}: the nominal-model VaR rather than the true VaR under the heavy-tailed data-generating process. This bias is structural and cannot be eliminated by increasing the simulation budget, as it arises from model misspecification rather than Monte Carlo noise.

Discrete moment matching, by contrast, does not rely on a fully specified parametric model. By restricting attention to distributions that satisfy a finite set of moment constraints, DMM explicitly acknowledges distributional ambiguity. As tail thickness increases, this ambiguity manifests as wider VaR intervals, reflecting increased uncertainty rather than spurious precision. Although the resulting estimates are less efficient and more conservative, they exhibit greater robustness to deviations from Gaussian tail behavior.

\subsection{Interpretation of Stability Diagnostics}
Stability diagnostics provide additional insight into the reliability of importance sampling under misspecification. Effective sample size and maximum weight concentration measure the extent to which the estimator is dominated by a small number of weighted observations. In the experiments considered here, these diagnostics remain within acceptable ranges under moderate tail thickness but deteriorate as tails become heavier.

Importantly, weight degeneracy in this setting does not necessarily imply estimator failure in a numerical sense; rather, it serves as an early warning that the change of measure is poorly aligned with the true tail geometry. In practice, such diagnostics should be monitored alongside point estimates, particularly when importance sampling is applied in environments where tail behavior is uncertain or subject to structural breaks.

\subsection{Implications for Market Risk Estimation}
The comparative results suggest that the choice between importance sampling and moment-matching approaches should be guided by the degree of confidence in the assumed tail model. When the nominal model is believed to be approximately correct and computational efficiency is paramount, importance sampling provides a powerful tool for rare-event estimation. Conversely, when tail behavior is uncertain or potentially heavy-tailed, moment-based approaches offer a more cautious alternative that trades efficiency for robustness.

In applied risk management settings, this distinction is particularly relevant. Regulatory and internal risk measures are often computed under simplifying assumptions that may not hold during periods of market stress. The present results indicate that reliance on variance-reduction techniques alone may lead to underestimation of extreme risk if model misspecification is not explicitly addressed.

\subsection{Limitations}
The present study adopts a deliberately controlled framework in order to
isolate the efficiency–robustness trade-off between importance sampling and
moment-based VaR bracketing under tail misspecification. As a result, several
dimensions of practical risk modeling are intentionally abstracted from. First, the analysis is restricted to a univariate return setting, and the behavior of both estimators in high-dimensional portfolio contexts may differ. Second, the discrete moment matching procedure depends on the choice of loss grid and the order of moments enforced; alternative grid constructions or higher-order moments may yield different bounds. Third, the importance sampling scheme considered here is limited to simple exponential tilting of a Gaussian model; more sophisticated proposal distributions may mitigate some of the observed bias under misspecification.

Finally, while discrete moment matching yields interval-valued VaR outputs
that naturally quantify distributional ambiguity, the present study does not
address how such intervals should be translated into single capital charges
in regulatory or operational settings. Developing principled decision rules
that map ambiguity intervals into actionable risk measures remains an
important direction for future research.

\section{Conclusion}
\label{sec:conclusion}

This paper studied the performance of two simulation-based approaches to Value-at-Risk estimation, importance sampling and discrete moment matching, under controlled tail misspecification. By separating the nominal model used for estimator construction from the true data-generating process used for evaluation, the analysis isolated the impact of heavy-tailed behavior on estimator bias, variability, and numerical stability.

The results demonstrate that importance sampling, while highly efficient under correct model specification, is sensitive to misspecification of tail behavior. In heavy-tailed environments, importance sampling converges efficiently to the VaR implied by the nominal Gaussian model, which can substantially underestimate the true risk. Stability diagnostics such as effective sample size and weight concentration provide useful indicators of this mismatch but do not eliminate the underlying structural bias.

Discrete moment matching exhibits complementary behavior. By restricting attention to distributions consistent with a finite set of moment constraints, DMM produces conservative VaR estimates that remain robust under heavy-tailed data-generating processes. This robustness comes at the cost of reduced efficiency and wider uncertainty intervals, reflecting the intrinsic ambiguity associated with limited distributional information.

Taken together, these findings underscore that variance reduction alone is not sufficient for reliable tail risk estimation in the presence of model uncertainty. The choice of VaR estimation methodology should therefore reflect not only computational considerations but also the degree of confidence in the assumed tail model. In settings where tail behavior is difficult to specify or subject to structural change, moment-based approaches may offer a safer alternative to aggressively optimized importance sampling schemes.

Future work may extend this analysis to higher-dimensional portfolio settings, alternative importance sampling proposals, and dynamic models of volatility and tail risk. Nonetheless, the present study provides a transparent and reproducible framework for understanding the efficiency–robustness trade-off inherent in simulation-based VaR estimation.

\bibliographystyle{plain}
\nocite{*}
\bibliography{references}

\end{document}